\documentclass[twocolumn,showpacs,preprintnumbers,amsmath,amssymb]{revtex4}
\usepackage{epsfig}
\usepackage{amsmath}
\usepackage{subfigure}
\usepackage{rotating}

\begin{document}

\title{Border trees of complex networks}

\author{Paulino R. Villas Boas}
\affiliation{Institute of Physics at S\~ao Carlos, University of
S\~ao Paulo, PO Box 369, S\~ao Carlos, S\~ao Paulo, 13560-970
Brazil}

\author{Francisco A. Rodrigues}
\affiliation{Institute of Physics at S\~ao Carlos, University of
S\~ao Paulo, PO Box 369, S\~ao Carlos, S\~ao Paulo, 13560-970
Brazil}

\author{Gonzalo Travieso}
\affiliation{Institute of Physics at S\~ao Carlos, University of
S\~ao Paulo, PO Box 369, S\~ao Carlos, S\~ao Paulo, 13560-970
Brazil}

\author{Luciano da Fontoura Costa}
\affiliation{Institute of Physics at S\~ao Carlos, University of
S\~ao Paulo, PO Box 369, S\~ao Carlos, S\~ao Paulo, 13560-970
Brazil}

\date{\today}

\begin{abstract}

The comprehensive characterization of the structure of complex
networks is essential to understand the dynamical processes which
guide their evolution. The discovery of the scale-free distribution
and the small world property of real networks were fundamental to
stimulate more realistic models and to understand some dynamical
processes such as network growth. However, properties related to the
network borders (nodes with degree equal to one), one of its most
fragile parts, remain little investigated and understood. The border
nodes may be involved in the evolution of structures such as
geographical networks. Here we analyze complex networks by looking for
border trees, which are defined as the subgraphs without cycles
connected to the remainder of the network (containing cycles) and
terminating into border nodes.  In addition to describing an algorithm
for identification of such tree subgraphs, we also consider a series
of their measurements, including their number of vertices, number of
leaves, and depth. We investigate the properties of border trees for
several theoretical models as well as real-world networks.

\end{abstract}

\maketitle

\section{\label{sec:intr}Introduction}

Complex networks are characterized by an uneven distribution of
connections which suggests that their growth is not defined by random
events. In this way, it is expected that some patterns emerge in their
structure which affect the dynamical aspects related to resilience,
transport and network maintenance. While such patterns, called network
motifs, have been largely characterized in last years
(e.g.~\cite{ShenOrr:2002,Villas-Boas07}), some of them remain
uncharacterized and their role in network function is not known. While
small network motifs are believed to be the building blocks of complex
networks~\cite{ShenOrr:2002}, larger motifs may emerge according to
different network needs and growth dynamics. For instance, $n$-chains
networks motifs~\cite{Villas-Boas07} can appear in order to provide
redundance of connections between two nodes, increasing the network
resilience to edge removal. Other motifs, such as border trees (as
well as other peripheric motifs), can be the result of the external
growth of the network, i.e., the network can evolve as a tree, where
each ``branch'' of nodes emerges from the main connected component to
the outside of the network.

In this work we provide a description of border tree motifs and
investigate the occurrence of such motifs in real-world networks as
well as networks generated by theoretical models.

\section{Border tree definition}

Although many measurements such as vertex degree, clustering
coefficient, shortest path length, betweenness centrality
(e.g.~\cite{Costa:2007:survey}), and many structures such as
motifs~\cite{ShenOrr:2002} and chains~\cite{Villas-Boas07} have been
defined, the characterization of complex networks is still
incomplete~\cite{Costa:2007:survey}, i.e.\ if we have a set of many
measurements we cannot fully recover the original corresponding
network. Therefore, new measurements or structures must be considered
for the study of complex networks according with the specific
needs. Here we introduce the concept of border trees in complex
network.

\begin{figure}
  \centerline{\includegraphics[width=0.6\linewidth]{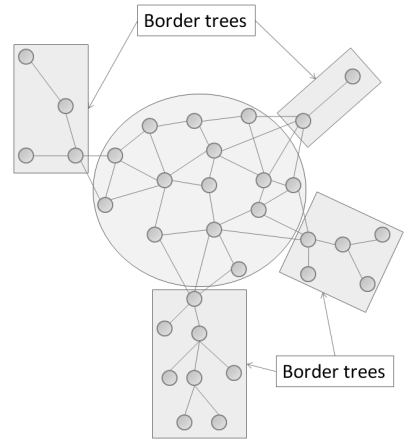}}
  \caption{Some examples of border trees of a small network.}
  \label{fig:btree}
\end{figure}

A border tree is a subgraph without cycles connected to the remainder
of the network (see Figure~\ref{fig:btree} for some examples). Its
root and leaves are, respectively, the vertex which belongs to a loop,
and the vertices with degree 1. Its depth is the largest distance
between its root and its leaves.

\section{Algorithm to find border trees}

Initially we find all vertices of degree 1 and create a tree for each
of them. For each tree, we verify whether the vertex at its top has
more than 2 neighbors, ignoring those at lower levels. If there is
more than 1, keep this tree in a waiting list. If there is just one,
add it to the tree and join any other trees in the auxiliary list
which has this vertex at its top. The algorithm ends when all trees
are in the auxiliary list, i.e. there is no way to join two trees, so
that searches for a vertex of the higher level fail. Note that the
tops of all the trees are vertices which belong to at least one loop.

\section{\label{sec:appl}Results and Discussion}

The models considered were the Erd\H{o}s and R\'enyi (ER) random
graph~\cite{Erdos-Renyi:1959}, the Watts and Strogatz (WS) small-world
model~\cite{Watts:1998}, Barab\'asi and Albert (BA) scale-free
model~\cite{Barabasi:99}, and a Geographical Network (GN) model as
described in~\cite{Villas-Boas07} where $N$ vertices are randomly
distributed inside a $L=\sqrt{N}$ length square and two vertices are
connected with probability $p\sim e^{-\lambda\,d}$, where $d$ is the
geographical distance between them and $\lambda$ is a model parameter
designed to generate the desired average vertex degree.

All analyzed models have $N=1000$ vertices and average degrees
$\langle k \rangle = $ 2, 4, and 6. The probability of connection in
the ER model is $\langle k \rangle/(N-1)$; the parameter $m$ is 1, 2,
and 3 for the BA model; $\kappa =$ 1, 2, and 3 and the probability for
the WS model is 0.2; and $\lambda=1.7$, $1.22$, and $0.97$, for the GN
model. Note that all parameters, except $q$ for the WS model, have
been chosen in order to guarantee average degree 2, 4, and 6 for all
models. A total of 100 realizations of each model were considered.

We considered 17 real-world networks divided into four classes:
social, information, technological, and biological networks. Their
descriptions and some of their most important measurements can be seen
in Table~\ref{tab:meas}.

\subsection{Basic Measurements}

Table~\ref{tab:meas} presents the description and the adopted
measurements of the considered networks, theoretical and
real-world. These measurements include the average vertex degree
$\langle k \rangle$, the average clustering coefficient $\langle c
\rangle$, and the average shortest path length
$\ell$~\cite{Costa:2007:survey}, and were obtained considering
unweighted networks. Those which were not originally of this type were
accordingly transformed to their unweighted counterpart by using the
threshold operation~\cite{Costa:2007:survey}. In the same way, the
directed networks were transformed into their undirected version by
using the symmetry operation~\cite{Costa:2007:survey} for the
calculation of the clustering coefficient. For the calculation of the
average shortest path length $\ell$, only the largest connected
component in the networks was considered.

\begin{table*}[h]
  \caption{Properties of the considered complex networks. $N$ is the number of
  vertices, $\langle k \rangle$ is the average degree, $\langle c
  \rangle$ is the average clustering coefficient, and $\ell$ is the
  average shortest path length.}
  \label{tab:meas}
  \begin{footnotesize}
\begin{tabular*}{0.99\textwidth}{@{\extracolsep{\fill}}l|l|p{0.4\linewidth}|c|c|r|r|r|r}
\hline
           & {\bf Networks} & {\bf Brief Description}& {\bf Directed} & {\bf Weighted} &  {\bf $N$} & {\bf $\langle k \rangle$} & {\bf $\langle c \rangle$} & {\bf $\ell$} \\
\hline
\:\:\:\begin{rotate}{90}\hbox{\hspace{-2.7cm}\it Models}\end{rotate}\: & ER  $\langle k\rangle = 2.03$ & Erd\H{o}s and R\'enyi random graph~\cite{Erdos-Renyi:1959} &         no &         no &      1 000 &       2.03 &      0.001 &       9.00 \\

           & \hspace{0.55cm}$\langle k\rangle = 4.01$ &            &         no &         no &      1 000 &       4.01 &      0.004 &       5.06 \\

           & \hspace{0.55cm}$\langle k\rangle = 6.01$ &            &         no &         no &      1 000 &       6.01 &      0.006 &       4.06 \\

           & WS $\langle k\rangle = 2$ & Watts and Strogatz small world model~\cite{Watts:1998} &         no &         no &      1 000 &       2.00 &      0.000 &      58.26 \\

           & \hspace{0.55cm}$\langle k\rangle = 4$ & with probability of rewiring 0.2           &         no &         no &      1 000 &       4.00 &      0.269 &       6.90 \\

           & \hspace{0.55cm}$\langle k\rangle = 6$ &            &         no &         no &      1 000 &       6.00 &      0.315 &       5.10 \\

           & BA $\langle k\rangle = 2$ & Barab\'asi and Albert scale-free model~\cite{Barabasi:99} &         no &         no &      1 000 &       2.00 &      0.000 &       6.92 \\

           & \hspace{0.55cm}$\langle k\rangle = 4$ &            &         no &         no &      1 000 &       4.00 &      0.031 &       4.02 \\

           & \hspace{0.55cm}$\langle k\rangle = 6$ &            &         no &         no &      1 000 &       6.00 &      0.037 &       3.45 \\

           & GN $\langle k\rangle = 2.08$ & Geographical Network model~\cite{Villas-Boas07} &         no &         no &      1 000 &       2.08 &      0.088 &      18.01 \\

           & \hspace{0.55cm}$\langle k\rangle = 3.97$ &            &         no &         no &      1 000 &       3.97 &      0.136 &       8.73 \\

           & \hspace{0.55cm}$\langle k\rangle = 6.18$ &            &         no &         no &      1 000 &       6.18 &      0.152 &       6.26 \\
\hline
\:\:\:\begin{rotate}{90}\hbox{\hspace{-1.7cm}\it Social}\end{rotate}\: & Astrophysics & Astrophysics collaboration network from 1995 to 1999~\cite{Newman-PNAS01} &         no &        yes &     16 706 &      14.52 &      0.639 &       4.80 \\

           & Netscience & Scientific collaboration of complex network researches compiled from~\cite{Newman:2003:survey,Boccaletti06} &         no &        yes &      1 461 &       3.75 &      0.638 &       5.82 \\

           &   Cond-mat & Condensed matter collaboration network from 1995 to 2005~\cite{Newman-PNAS01} &         no &        yes &     40 421 &       8.69 &      0.636 &       5.50 \\

           & High-energy theory & High-energy theory collaboration network from 1995 to 1999~\cite{Newman00:PRE64:I,Newman00:PRE64:II} &         no &        yes &      8 361 &       3.77 &      0.442 &       7.03 \\
\hline
\:\:\:\begin{rotate}{90}\hbox{\hspace{-2.2cm}\it Information}\end{rotate}\: & Roget network & Roget's thesaurus network~\cite{Roget82,pajek-data} &        yes &         no &      1 022 &       4.97 &      0.150 &       4.90 \\

           &    Wordnet & Semantic network~\cite{pajek-data} &        yes &         no &     82 670 &       1.60 &      0.027 &       9.15 \\

           &        WWW & World Wide Web, network of web pages~\cite{Albert99:Nature,CCNR} &        yes &         no &    325 729 &       4.51 &      0.235 &      11.27 \\

           & David Copperfield & Word adjacency network from the book David Copperfield by Charles Dickens~\cite{Antiqueira2006,Antiqueira2007} &        yes &        yes &     11 378 &      10.05 &      0.218 &       3.60 \\

           & Night and Day & Word adjacency network from the book Night and Day by Virginia Woolf~\cite{Antiqueira2006,Antiqueira2007} &        yes &        yes &      7 959 &       7.83 &      0.145 &       3.81 \\

           & On the origin of species & Word adjacency network from the book On the origin of species by Charles Darwin~\cite{Antiqueira2006,Antiqueira2007} &        yes &        yes &      6 973 &       9.57 &      0.181 &       3.87 \\
\hline
\:\:\:\begin{rotate}{90}\hbox{\hspace{-1.5cm}\it Technological}\end{rotate}\: &   Internet & Autonomous system network is a collection of IP networks and routers~\cite{Newman:data} &         no &         no &     22 963 &       4.22 &      0.230 &       3.84 \\

           &    Airport & US airlines transportation network is formed by airports in 1997 connected by flights~\cite{pajek-data} &         no &        yes &        332 &      12.81 &      0.626 &       2.74 \\

           & Power grid & Western states power grid network~\cite{Watts:1998} &         no &         no &      4 941 &       2.67 &      0.080 &      18.99 \\

\hline
\:\:\:\begin{rotate}{90}\hbox{\hspace{-1.5cm}\it Biological}\end{rotate}\: &   Food web & Food web of Florida Bay Trophic~\cite{pajek-data} &        yes &        yes &        128 &      16.70 &      0.335 &       2.41 \\

           & \emph{C.\ elegans} & Neural network of \emph{Caenorhabditis elegans}~\cite{White86,Watts:1998} &        yes &        yes &        297 &       7.95 &      0.293 &       3.99 \\

           &    \emph{E.\ coli} & Transcriptional regulation network of the \emph{Escherichia coli}~\cite{ShenOrr:2002} &        yes &        yes &        423 &       1.23 &      0.085 &       1.36 \\

           & \emph{S.\ cerevisiae} & Protein-protein interaction network of \emph{Saccharomyces cerevisiae}~\cite{Jeong01} &         no &         no &      2 708 &       5.26 &      0.188 &       4.74 \\
\hline
\end{tabular*}
  \end{footnotesize}
\end{table*}

\subsection{Statistics of border trees}

Table~\ref{tab:btrees} presents the average, mode, and the maximum of
the number of nodes, the depth, the number of children per vertex and
the number of leaves per tree for each of the theoretical and
real-world networks. In the former case, the measurements refer to the
average of 100 realizations of each configuration.

\begin{table*}
  \caption{Statistics of border trees in networks.} \label{tab:btrees}
  \begin{center}
  \begin{footnotesize}
\begin{tabular*}{0.9\textwidth}{@{\extracolsep{\fill}}l|l|rrr|rrr|rrr|rrr}
\hline
           &            & \multicolumn{ 3}{ c|}{{\bf Number of nodes}} &    \multicolumn{ 3}{c|}{{\bf Depth}} & \multicolumn{ 3}{c|}{{\bf Number of children}} & \multicolumn{ 3}{c}{{\bf Number of leaves}} \\
\cline{3-14}
           & {\bf Network} & {\bf Mean} & {\bf Mode} & {\bf Max } & {\bf Mean} & {\bf Mode} & {\bf Max } & {\bf Mean} & {\bf Mode} & {\bf Max } & {\bf Mean} & {\bf Mode} & {\bf Max } \\
\hline
\:\:\:\begin{rotate}{90}\hbox{\hspace{-2.3cm}\it Models}\end{rotate}\: & ER  $\langle k\rangle = 2.03$ &       3.06 &          2 &         25 &       1.61 &          1 &         10 &       1.21 &          1 &          5 &       1.40 &          1 &         10 \\

           & \hspace{0.55cm}$\langle k\rangle = 4.01$ &       2.12 &          2 &          7 &       1.08 &          1 &          5 &       1.04 &          1 &          3 &       1.04 &          1 &          4 \\

           & \hspace{0.55cm}$\langle k\rangle = 6.01$ &       2.03 &          2 &          4 &       1.02 &          1 &          3 &       1.01 &          1 &          2 &       1.01 &          1 &          2 \\

           & WS $\langle k\rangle = 2$ &      47.84 &          2 &        987 &      14.37 &          1 &        167 &       1.10 &          1 &          2 &       8.15 &          1 &        165 \\

           & \hspace{0.55cm}$\langle k\rangle = 4$ &         -- &         -- &         -- &         -- &         -- &         -- &         -- &         -- &         -- &         -- &         -- &         -- \\

           & \hspace{0.55cm}$\langle k\rangle = 6$ &         -- &         -- &         -- &         -- &         -- &         -- &         -- &         -- &         -- &         -- &         -- &         -- \\

           & BA $\langle k\rangle = 2$ &       1000 &       1000 &       1000 &       8.93 &          9 &         12 &       3.01 &          3 &          3 &     667.88 &        663 &        697 \\

           & \hspace{0.55cm}$\langle k\rangle = 4$ &         -- &         -- &         -- &         -- &         -- &         -- &         -- &         -- &         -- &         -- &         -- &         -- \\

           & \hspace{0.55cm}$\langle k\rangle = 6$ &         -- &         -- &         -- &         -- &         -- &         -- &         -- &         -- &         -- &         -- &         -- &         -- \\

           & GN $\langle k\rangle = 2.08$ &       3.26 &          2 &         38 &       1.70 &          1 &         15 &       1.23 &          1 &          4 &       1.47 &          1 &         15 \\

           & \hspace{0.55cm}$\langle k\rangle = 3.97$ &       2.27 &          2 &         10 &       1.18 &          1 &          7 &       1.06 &          1 &          3 &       1.09 &          1 &          5 \\

           & \hspace{0.55cm}$\langle k\rangle = 6.18$ &       2.10 &          2 &          6 &       1.07 &          1 &          4 &       1.03 &          1 &          3 &       1.03 &          1 &          3 \\
\hline
\:\:\:\begin{rotate}{90}\hbox{\hspace{-.9cm}\it Social}\end{rotate}\: & Astrophysics &       2.35 &          2 &          8 &       1.06 &          1 &          3 &       1.27 &          1 &          6 &       1.29 &          1 &          6 \\

           & Netscience &       2.30 &          2 &          6 &       1.01 &          1 &          2 &       1.26 &          1 &          3 &       1.27 &          1 &          3 \\

           &   Cond-mat &       2.43 &          2 &          9 &       1.06 &          1 &          3 &       1.34 &          1 &          8 &       1.37 &          1 &          8 \\

           & High-energy theory &       2.55 &          2 &         10 &       1.15 &          1 &          6 &       1.34 &          1 &          4 &       1.40 &          1 &          7 \\
\hline
\:\:\:\begin{rotate}{90}\hbox{\hspace{-1.6cm}\it Information}\end{rotate}\: & Roget network &       2.40 &          2 &          5 &       1.29 &          1 &          3 &       1.08 &          1 &          2 &       1.12 &          1 &          2 \\

           &    Wordnet &       5.41 &          2 &        211 &       1.25 &          1 &          7 &       2.96 &          1 &         84 &       4.06 &          1 &        208 \\

           &        WWW &      10.73 &          2 &       5329 &       1.13 &          1 &         21 &       6.83 &          1 &       1430 &       9.48 &          1 &       5324 \\

           & David Copperfield &       2.24 &          2 &          4 &       1.00 &          1 &          1 &       1.24 &          1 &          3 &       1.24 &          1 &          3 \\

           & Night and Day &       2.38 &          2 &          4 &       1.00 &          1 &          1 &       1.38 &          1 &          3 &       1.38 &          1 &          3 \\

           & On the origin of species &       2.09 &          2 &          3 &       1.00 &          1 &          1 &       1.09 &          1 &          2 &       1.09 &          1 &          2 \\
\hline
\:\:\:\begin{rotate}{90}\hbox{\hspace{-1.5cm}\it Technological}\end{rotate}\: &   Internet &       5.67 &          2 &        329 &       1.07 &          1 &          3 &       3.73 &          1 &        173 &       4.58 &          1 &        327 \\

           &    Airport &       3.12 &          2 &         13 &       1.00 &          1 &          1 &       2.12 &          1 &         12 &       2.12 &          1 &         12 \\

           & Power grid &       2.97 &          2 &         21 &       1.39 &          1 &          7 &       1.31 &          1 &          9 &       1.52 &          1 &         12 \\

           &            &            &            &            &            &            &            &            &            &            &            &            &            \\

           &            &            &            &            &            &            &            &            &            &            &            &            &            \\
\hline
\:\:\:\begin{rotate}{90}\hbox{\hspace{-1.1cm}\it Biological}\end{rotate}\: &   Food web &         -- &         -- &         -- &         -- &         -- &         -- &         -- &         -- &         -- &         -- &         -- &         -- \\

           & \emph{C.\ elegans} &       6.00 &          2 &         11 &       1.00 &          1 &          1 &       5.00 &          1 &         10 &       5.00 &          1 &         10 \\

           &    \emph{E.\ coli} &       6.14 &          3 &         24 &       1.30 &          1 &          3 &       3.82 &          1 &         21 &       4.61 &          2 &         21 \\

           & \emph{S.\ cerevisiae} &       2.98 &          2 &         22 &       1.21 &          1 &          3 &       1.61 &          1 &         11 &       1.75 &          1 &         18 \\
\hline
\end{tabular*}
  \end{footnotesize}
  \end{center}
\end{table*}

For all considered networks, the border trees have typically 2
vertices (one leave and one parent --- a vertex which belongs to the
remainder of the network) and depth 1. The exceptions are all models
with average degree 2 and Wordnet, WWW, Internet, Airport, Power grid,
Food web, \emph{C.\ elegans}, \emph{E.\ coli}, \emph{and S.\
cerevisiae} networks.

Interesting results concern the WS and BA models with average degree
2, WWW, Food web, \emph{C.\ elegans}, \emph{E.\ coli}, and \emph{S.\
cerevisiae} networks. The WS model with average degree 2 has the
longest tree depth because of the formation of linear chains of
vertices after the rewiring process of the initial configuration (ring
of vertices).  The BA model with average degree 2 has a tree-like
structure and, therefore, presents the largest values for all
measurements, except the average and maximum depth and number of
children, and the maximum number of leaves. The WWW resulted with the
greatest number of vertices, the greatest number of children, and the
greatest number of leaves in a tree, and also has large averages, but
the most frequent tree has 2 vertices (one leave and one parent). On
the other hand, the Food web does not present trees. This kind of
network is essentially compounded by loops, since every living
creature is connected to the decomposers.

\section{\label{sec:conc}Conclusions}

This work has introduced the concept of border tree and presented a
simple and effective algorithm for their identification.  Statistics
of the presence of such motifs in several real-world and theoretical
networks were obtained and shown to provide valuable information
regarding the overall structure of the analysed networks.  Overall,
markedly distinct statistics of border trees were obtained for the
considered models, which corroborates the potential of such
measurements for the discrimination and identification of networks.
Unlike what was recently observed for chain
motifs~\cite{Villas-Boas07}, border trees were found for both
theoretical and real-world networks.  Among the the former, we
obtained the largest tree for the BA with average degree equal to two,
while the WS models exhibited the longest depths.  In the case of the
real-world networks, the WWW presented the largest overall
measurements, suggesting that this network involves a larger number of
significative trees around its borders, possibly corresponding to the
more recently included nodes.  The Internet and power-grid network (a
geographical structure) presented similar properties, though
exhibiting shortest depths.  Among the biological networks, the
neuronal network of \emph{C. elegans} and the transcription network of
\emph{E. coli} presented the largest number of nodes belonging to border 
trees.

\begin{acknowledgments}

The authors thank Lucas Antiqueira for providing the book networks.
Luciano da F. Costa thanks CNPq (301303/06-1) and FAPESP (05/00587-5).
Francisco A. Rodrigues is grateful to FAPESP (04/00492-1) and Paulino
R. Villas Boas is grateful to CNPq (141390/2004-2).

\end{acknowledgments}

\bibliography{paper}

\end{document}